# Suppression of hypersynchronous network activity in cultured cortical neurons using an ultrasoft silicone scaffold

Takuma Sumi[a], Hideaki Yamamoto*[b], and Ayumi Hirano-Iwata[ab]

[a]*Research Institute of Electrical Communication, Tohoku University, 2-1-1 Katahira, Aoba-ku, Sendai 980-8577, Japan. E-mail: hideaki.yamamoto.e3@tohoku.ac.jp*

[b]*WPI-Advanced Institute for Materials Research (WPI-AIMR), Tohoku University, 2-1-1 Katahira, Aoba-ku, Sendai 980-8577, Japan*

**Abstract**

The spontaneous activity pattern of cortical neurons in dissociated culture is characterized by burst firing that is highly synchronized among a wide population of cells. The degree of synchrony, however, is excessively higher than that in cortical tissues. Here, we employed polydimethylsiloxane (PDMS) elastomers to establish a novel system for culturing neurons on a scaffold with an elastic modulus resembling brain tissue, and investigated the effect of the scaffold's elasticity on network activity patterns in cultured rat cortical neurons. Using whole-cell patch clamp to assess the scaffold effect on the development of synaptic connections, we found that the amplitude of excitatory postsynaptic current, as well as the frequency of spontaneous transmissions, was reduced in neuronal networks grown on an ultrasoft PDMS with an elastic modulus of 0.5 kPa. Furthermore, the ultrasoft scaffold was found to suppress neural correlations in the spontaneous activity of the cultured neuronal network. The dose of GsMTx-4, an antagonist of stretch-activated cation channels (SACs), required to reduce the generation of the events below 1.0 event/min on PDMS substrates was lower than that for neurons on a glass substrate. This suggests that the difference in the baseline level of SAC activation is a molecular mechanism underlying the alteration in neuronal network activity depending on scaffold stiffness. Our results demonstrate the potential application of PDMS with biomimetic elasticity as cell-culture scaffold for bridging the *in vivo-in vitro* gap in neuronal systems.

**Main text**

**1. Introduction**

*In vitro* modelling of *in vivo* multicellular functions is essential in biology and medicine not only for basic studies but also for applied research, such as the screening of candidate molecules in drug development.[1,2] In fields such as cardiology and oncology, cultured-cell models have been established and are used in disease modelling and toxicity assays.[1,3] However, in neuroscience, cortical and hippocampal neurons in dissociated culture generate a non-physiological activity characterized by globally synchronized burst firing, often referred to as 'network bursts'.[4-7] This activity pattern is significantly different from that observed in an animals' cortex or hippocampus, which is highly complex both spatially and temporally.[8,9] Such complexity in neural activity is important, as it underlies the computational capacity of the neuronal networks.[10,11]

Several approaches have been taken to suppress the globally synchronized bursting in cultured neuronal networks. For instance, it has been shown that the synchronized bursts are inhibited and the complexity in the spontaneous activity is upregulated by growing cultured neurons on micropatterned surfaces to induce a network architecture such as those observed in the *in vivo* networks.[12] The role of external inputs in shaping the spontaneous dynamics of the cultured neural networks has also been investigated both experimentally and computationally, showing that chronic application of external stimulus that resembles thalamic input decorrelates cortical neuronal network activity.[13-15] Furthermore, pharmacological blockade of an AMPA-type glutamate receptor with CNQX at a dose below its $IC_{50}$ reduces the spatial extent of the burst spreading,[5] possibly through a reduction in the excitatory synaptic strength that is excessively strong in cultured neurons as compared to the *in vivo* cortex.[16-18]

Another major difference between the *in vitro* and *in vivo* neuronal networks is the mechanical property of their scaffolds. Cultured neurons are usually grown on a polystyrene or glass substrate, whose elastic moduli, *E*, are in the order of GPa.[19,20] In contrast, the brain is the softest tissue in an animals' body, with an *E* below 1 kPa.[21] Several studies on non-neuronal cells have pointed to the importance of culturing cells on a scaffold with biomimetic elasticity. For instance, mesenchymal stem cells commit to the lineage specified by scaffold elasticity.[22] Furthermore, the expression of chondrocyte phenotype is stabilized when cultured on a scaffold with an *E* of 5.4 kPa, similar to that of the *in vivo* environment.[23] Based on these observations, we hypothesized that the non-physiological synchronized bursting in cultured neuronal networks could be suppressed by growing neurons on a biomimetic scaffold.

In this work, we established a biomimetic culture platform using polydimethylsiloxane (PDMS) that is as soft as brain tissue (i.e. *E* ~ 0.5 kPa). PDMS is a well-established biocompatible material, whose elasticity can be tuned in a wide range, from ~0.1 kPa to tens of MPa by choosing the precursors and changing their mixing ratio.[24,25] It also offers several advantages over more commonly used materials (e.g. polyacrylamide), such as being compatible with surface modification techniques, being electrically insulating, and having a long shelf life.[26] Primary rat cortical neurons were cultured on the PDMS substrate, and the effect of the scaffold's stiffness on synaptic strength and the complexity of the neuronal network activity was assessed using whole-cell patch-clamp recording and fluorescent calcium imaging, respectively. We show that the excitatory synapses are weakened on the softer substrates and that the neuronal correlation in spontaneous network activity is significantly reduced on the PDMS substrate with an *E* ~ 0.5 kPa. The underlying molecular mechanism responsible for the stiffness-dependent modulation on spontaneous network activity is pharmacologically explored by blocking stretch-activated cation channels (SACs).

## 2. Experimental

### 2.1 Mechanical characterization of the PDMS

PDMS was prepared using Sylgard 184 (Dow Corning; mixing ratio = 50:1) and Sylgard 527 (Dow Corning; mixing ratio = 5:4). For each PDMS, 200 g of the mixtures were poured in a glass petri dish (diameter, 90 mm; height, 60 mm), degassed in a vacuum chamber, and cured in an oven (AS-ONE SONW-450S) for two days at 80 °C.

The elastic modulus of the PDMS was determined by the spherical indentation method (Fig. 1a) following Zhang et al.[27,28] Briefly, a chromium steel ball of 3.175-mm radius ($R$) was attached onto the load cell of the Instron 5943 Universal Testing System. The depth ($\delta$)-indentation load ($P$) curves were measured (Fig. 1b), and the elastic moduli, $E$, were determined by fitting the load curves to the following equation:

$$P = \frac{16}{9} E \sqrt{R\delta} \, \delta \left(1 - 0.15 \frac{\delta}{R}\right). \tag{1}$$

### 2.2 PDMS substrates for neuronal culture

Glass coverslips (Matsunami C018001; diameter, 18 mm; thickness 0.17 mm) were first cleaned by sonication in 99.5% ethanol and rinsed two times in Milli-Q grade water. After a thorough mixing of the two PDMS components and subsequent degassing, 100 μL of the mixture was drop casted on the coverslip. PDMS was then cured in an oven for 11 h at 80 °C.

### 2.3 Contact angle measurement

The hydrophilicity of the surfaces was characterized by measuring the water contact angle. Using the LSE-B100 equipment (NiCK Corporation, Japan), a 0.5-μL water droplet was dropped onto the substrate and was imaged from the side. The contact angle of the droplet was measured using the i2win software (NiCK Corporation, Japan). Three samples were prepared for each condition, and measurements were performed at three different positions for each sample.

## 2.4 Cell culture

For cell culturing, the PDMS substrate was first treated in air plasma (Yamato PM-100) for 10 s and was sterilized under UV light for 60 min. The surface of the PDMS was then coated with poly-D-lysine (PDL; Sigma P-0899) by floating the sample upside-down on a phosphate-buffered saline (Gibco 14190-144) containing 50 μg/mL PDL overnight. The sample was then rinsed two times in sterilized water and dried in air inside a laminar flow hood. One day prior to cell plating, the sample was immersed in the plating medium [minimum essential medium (Gibco 11095-080) + 5% foetal bovine serum + 0.6% D-glucose] and stored in a $CO_2$ incubator (37 ºC). Glass coverslips without the PDMS layer were used in control experiments. These were prepared by cleaning coverslips in ethanol and water, treating the surface with air plasma (60 s), UV-sterilization (60 min), and subsequent coating with PDL (overnight).

Rat cortical neurons from 18-d old embryos were used in our experiments. All procedures were approved by the Center for Laboratory Animal Research, Tohoku University (approval number: 2017AmA-001-1). After dissection of the cortical tissues and cell dispersion, the cells were plated on the samples immersed in the plating medium. After a 3 h incubation, the medium was changed to Neurobasal medium [Neurobasal (Gibco 21103-049) + 2% B-27 supplement (Gibco 17504-044) + 1% GlutaMAX-I (Gibco 3505-061)]. Half of the medium was replaced with fresh Neurobasal medium at 4 and 8 days of the culture.

**2.5 Electrophysiology**

Whole-cell patch-clamp recordings (HEKA EPC-10) were performed on neurons at 14−18 DIV under the voltage-clamp mode (holding potential, -70 mV). Signals were sampled at 20 kHz and filtered with 10 kHz and 2.9 kHz Bessel filters. Recordings were performed at room temperature. The intracellular solution contained: 146.3 mM KCl, 0.6 mM $MgCl_2$, 4 mM ATP-Mg, 0.3 mM GTP-Na, 5 U/mL creatine phosphokinase, 12 mM phosphocreatine, 1 mM EGTA, and 17.8 mM HEPES (pH 7.4). The extracellular solution for the recording contained: 140 mM NaCl, 2.4 mM KCl, 10 mM HEPES, 10 mM glucose, 2 mM $CaCl_2$, and 1 mM $MgCl_2$ (pH 7.4).[18] The $GABA_A$ receptor antagonist, bicuculline (Sigma 14343; 10 μM), was added to the extracellular solution to block inhibitory synaptic transmission. The membrane resistance was ~30 MΩ, and the synaptic currents with amplitude of 10−150 pA were analysed using a custom code written in MATLAB (Mathworks).

**2.6 Fluorescent calcium imaging**

Cultured neurons were loaded with a fluorescence calcium indicator Cal-520 AM (AAT Bioquest).[12] The cells were first rinsed in HEPES-buffered saline (HBS) containing 128 mM NaCl, 4 mM KCl, 1 mM $CaCl_2$, 1 mM $MgCl_2$, 10 mM D-glucose, 10 mM HEPES, and 45 mM sucrose, and subsequently incubating in HBS containing 2 μM Cal-520 AM for 30 min at 37 °C. The cells were then rinsed in fresh HBS and were imaged on an inverted microscope (IX83, Olympus) equipped with a 20× objective lens (numerical aperture, 0.70), a light-emitting diode light source (Lambda HPX, Sutter Instrument), a scientific complementary metal-oxide semiconductor camera (Zyla 4.2, Andor), and an incubation chamber (Tokai Hit). All recordings were performed at 14−18 DIV, while incubating in HBS at 37 °C. In some experiments, GsMTx-4 (Peptide Institute 4393-s) was added to the HBS to inhibit SACs.[29] Each recording was performed for 10 min at a frame rate of 10 Hz.

## 3. Results and discussion

### 3.1 Material properties of silicone scaffolds

The elastic scaffolds for neuronal culture were prepared with two types of PDMS, i.e. Sylgard 184 mixed at a ratio of 50:1 (hereafter referred to as 'soft') and Sylgard 527 mixed at a ratio of 5:4 (hereafter referred to as 'ultrasoft'). We first prepared the PDMS in glass petri dishes and determined their elastic moduli by the spherical indentation method[27,28] (Fig. 1). The elastic moduli of soft and ultrasoft PDMS were determined to be 13.6 ± 1.1 kPa (mean ± S.D.; $n = 4$) and 0.5 ± 0.03 kPa ($n = 5$), respectively (Fig. 1c). The values are in good agreement with previous studies,[24,27] and the elastic modulus of the ultrasoft PDMS was nearly equal to that of brain tissue.[21]

We next evaluated the wettability of the PDMS surface by measuring water contact angles. Neurons require the scaffold surface to be coated with cationic molecules, such as PDL. However, the strong hydrophobicity of as-prepared PDMS prevents the molecules from stably adsorbing on the surface.[30] Therefore, the samples were exposed to air plasma for a designated amount of time, which hydrophilizes the PDMS surface by substituting methyl groups with hydroxyl groups.[31] The changes in water contact angle $\theta$ of the soft and ultrasoft PDMS upon the plasma treatment are shown in Fig. 2a. Prior to the plasma treatment, the PDMS surface was hydrophobic, and $\theta$ were measured to be 132.3 and 128.0 for the soft and ultrasoft PDMS, respectively. The hydrophilicities of samples increased with the plasma exposure time. The hydrophilized surface was finally coated with PDL, and rat cortical neurons were cultured on the substrates. Plain glass coverslips coated with PDL were used as controls. For the cell-culture experiment, samples exposed to the plasma for 10 s were used in order to minimize the effect of surface vitrification and cracking.[31,32] As shown in Fig. 2a, $\theta$ for the soft and ultrasoft PDMS immediately after the 10 s plasma treatment were significantly different. The values of $\theta$ for the two scaffolds were found to converge after the PDL and the subsequent immersion in the neuronal plating medium (Fig. 2b). Representative micrographs of the rat cortical neurons cultured on the soft and ultrasoft PDMS are shown in Figs. 2c–e. The cell bodies of the neurons were well spread, and the neurites uniformly covered the entire surface. In order to compensate for the difference in cell affinity between glass and PDMS, initial plating density was increased 1.5-fold for the two PDMS scaffolds to achieve a constant attachment density of ~950 cells/mm$^2$ (Fig. 2f).

**3.2 Reduction of excitatory synaptic currents on ultrasoft scaffolds**

Previous work has shown that the amplitude of excitatory postsynaptic current (EPSC) in hippocampal neurons cultured on Sylgard 184 with $E$ = 457 kPa was significantly higher than that of neurons on Sylgard 184 with $E$ = 46 kPa.[27] To investigate whether a further reduction of substrate stiffness to mimic that of the brain tissue ($E \sim 0.5$ kPa) influences the synaptic strengths, we compared the amplitude and frequency of spontaneous EPSC (sEPSC) in neuronal networks grown on the soft ($E$ = 14 kPa) and ultrasoft ($E$ = 0.5 kPa) PDMS. sEPSC was recorded from cultured cortical neurons at 14−18 DIV under whole-cell patch clamp. To inhibit spontaneous inhibitory transmissions, a GABA$_A$ receptor blocker, bicuculline (10 μM), was added to the extracellular solution during recording

Representative traces from neurons cultured on glass, soft PDMS, and ultrasoft PDMS are shown in Figs. 3a–c, respectively. The amplitude of sEPSC observed in the neurons on soft substrates was 15% lower than those on glass substrates [soft: 23.5 ± 4.1 pA (mean ± S.D.; $n$ =

13), glass: 27.8 ± 7.0 pA (*n* = 11)]. sEPSC amplitude in neurons on ultrasoft substrates was further reduced from those on soft substrates and was approximately 30% lower than those on glass substrates [ultrasoft: 20.4 ± 2.3 pA (*n* = 12)]. In addition, the frequency of sEPSC from the neurons on soft and ultrasoft substrates was significantly lower than that on glass substrates (ultrasoft: 8.7 ± 3.3 Hz, soft: 9.7 ± 3.3 Hz, glass: 13.3 ± 5.6 Hz). These data are summarized in Figs. 3d and 3e. These results indicate that ultrasoft substrates that resemble the elastic moduli of brain tissues suppress the excitatory synaptic strength in cultured cortical neurons. The molecular mechanisms underlying the observations are further investigated and discussed in section 3.4.

## 3.3 Suppression of neural synchrony on ultrasoft scaffolds

Next, fluorescence calcium imaging was used to quantify the difference in the spontaneous firing patterns of neuronal networks on respective substrates. Representative traces of relative fluorescence intensity ($\Delta F/F_o$) from single neurons are shown in Figs. 4a–c. On the glass surface, the peak amplitude of the calcium transients was 0.42 ± 0.01, and the rate was 9.7 ± 0.2 events/min (*n* = 500). Both the peak amplitude and the event rate were significantly reduced on the soft PDMS (0.37 ± 0.01 and 7.1 ± 0.3 events/min, respectively; *n* = 500). On the ultrasoft substrates, both the amplitude and rate were further reduced as compared to the soft substrate and the control (0.27 ± 0.01 and 5.5 ± 0.2 events/min, respectively; *n* = 500). The reduction is likely to be caused by the reduction in the excitatory synaptic strength. These data are summarized in Figs. 4d and 4e.

In order to analyse the degree of neural correlations in the spontaneous activity, we evaluated the correlation coefficient, $r_{ij}$, between neurons *i* and *j*, as:

$$r_{ij} = \frac{\sum_t (f_i(t) - \bar{f}_i)(f_j(t) - \bar{f}_j)}{\sqrt{\sum_t (f_i(t) - \bar{f}_i)^2} \sqrt{\sum_t (f_j(t) - \bar{f}_j)^2}}, \tag{2}$$

where $f_i(t)$ is the relative fluorescence intensity of cell $i$ at time $t$, and the overline represents time average. Then, we compared their mean, $\bar{r} = (\sum_{i,j(i \neq j)} r_{ij})/N^2$, where $N$ (= 50) is the total number of analysed neurons on respective substrates. Although no significant difference in $\bar{r}$ was observed between glass and soft substrates, the value was significantly lower in the neuronal network grown on the ultrasoft scaffold (Fig. 4f). These results show that excessive neural synchronization was suppressed by reducing the scaffold stiffness to 0.5 kPa.

The results obtained in this work are in agreement with the previous study, which showed that a stiff PDMS substrate with $E$ = 457 kPa increased hippocampal neuronal network activity as compared to a PDMS substrate with $E$ = 46 kPa.[27] However, no discernible change in network synchrony was observed within the range of the elasticities investigated by the previous study. In the present study, we found that the non-physiological bursting activity is suppressed, and the mean correlation coefficient significantly decreases when the elastic modulus of the scaffold is further reduced to 0.5 kPa. Thus, Sylgard 527 is a promising scaffold for suppressing the hypersynchrony in neuronal culture.

**3.4 Molecular mechanism of the scaffold effect**

The above results show that the ultrasoft scaffold weakens the excitatory synaptic strength and reduces the synchrony in the neuronal network activity. We hypothesized that SACs, whose activity is downregulated on softer substrates,[33] would be the underlying molecular mechanism and investigated the effect of its pharmacological blockade on the neuronal network activity.

GsMTx-4 is a selective antagonist for SACs with an equilibrium constant of approximately 500 nM.[29,34] We first investigated the effect of reducing SAC activity in neurons on glass substrates. Bath application of GsMTx-4 at a concentration of 250 nM was found to reduce the peak amplitude and the rate of spontaneous calcium transients [0.30 ± 0.01 and 4.2 ± 0.1 events/min (mean ± S.E.M.), respectively; Fig. 5]. When GsMTx-4 was applied at a higher concentration of 500 nM, the rate was further reduced to 0.24 ± 0.01 events/min (Fig. 5b), while the peak amplitude did not significantly vary from the value observed at 250 nM (Fig. 5a). These results indicate that the fraction of active SACs in the neuronal plasma membrane plays a key role in the generation of spontaneous bursting events and the size of individual events.

We next examined the impact of GsMTx-4 application on cortical neurons grown on the PDMS substrates. Application of GsMTx-4 at a concentration of 250 nM reduced the rate of spontaneous calcium transients down to 0.62 ± 0.06 and 0.42 ± 0.02 events/min on the soft and ultrasoft substrates, respectively (Fig. 5b). Therefore, the dose of GsMTx-4 required to reduce the spontaneous occurrence of the calcium transients below 1.0 event/min was lower than that for the neurons on the glass substrate. This suggests that the difference in the baseline level of SAC activation is a molecular mechanism that contributes to the alteration in neuronal network activity depending on scaffold stiffness.

Penn *et al.*[35] previously showed that synchronized network activity in cultured hippocampal neurons decreased with extracellular calcium concentration, which was discussed to be caused by a reduction in presynaptic vesicle release probability. Considering that SACs permeate calcium ions,[36] the decrease in SAC activation could underlie the reduction in sEPSC amplitude and frequency, and neuronal synchrony on ultrasoft substrates.[35,37] Another possibility is that the influx of sodium ions through SACs[36] could directly enhance neuronal excitability independent of the modulation of synaptic strength (e.g. through facilitation of action potential generation). Finally, a mechanism independent of SACs could also have a role. A recent study reported that stiff substrates increase the number of synapses and reduce voltage-dependent $Mg^{2+}$ blockade in N-methyl-D-aspartate receptors, which lead to higher postsynaptic activity in cultured hippocampal neurons.[38] Figure 6 summarizes the above discussion concerning the underlying molecular mechanisms for the suppression of hypersynchrony on the ultrasoft substrate.

## 4. Conclusions

We established a protocol for culturing primary cortical neurons on an ultrasoft PDMS gel that mimics the elasticity of brain tissues and investigated the impact of the biomimetic scaffold on synaptic strength and spontaneous activity patterns. Our study showed that the ultrasoft substrate reduces the amplitude of sEPSCs (Fig. 3) that are excessively strong in the *in vitro* cultures. This led to significant reduction in the peak fluorescence amplitude and event rate of spontaneous network bursts on the ultrasoft substrate as compared to the glass substrate (Fig. 4). No significant difference in the correlation of neuronal network activity was observed on the scaffolds with $E > 13.5$ kPa. In contrast, this value was significantly lower for the neuronal network grown on the scaffold with $E = 0.5$ kPa (Fig. 4f), a stiffness similar to that of brain tissue. This is the first evidence that the ultrasoft scaffold with biomimetic elasticity effectively suppresses the hypersynchrony in the spontaneous network activity. A difference in the baseline activation of SACs underlie these stiffness-dependent changes in synaptic transmission and neuronal network activity.

Understanding of cellular mechanosensitivity has advanced rapidly since Engler *et al*.[22] found in 2006 that mesenchymal stem cells commit to the lineage specified by scaffold elasticity. The ultrasoft PDMS scaffold offers a mechanically biomimetic culture platform that is beneficial in suppressing the synchronous bursting in neuronal cultures. Moreover, it is a useful platform to study the influence of mechanical cues on neuronal network development. Further work is necessary to fully suppress the synchronized bursting in neuronal cultures. This could be accomplished by integrating cell micropatterning technology with ultrasoft scaffolds or by adding external noise to fill in for functional interactions between brain regions.[12,13]

**Conflicts of interest**

There are no conflicts to declare.

**Acknowledgements**

We acknowledge Prof. Hisashi Kino and Prof. Tetsu Tanaka of Tohoku University for the mechanical analysis of PDMS. This work was supported by the Japan Society for the Promotion of Science (Kakenhi Grant No. 18H03325) and by the Japan Science and Technology Agency (PRESTO: JPMJPR18MB and CREST: JPMJCR14F3).

**Figures**

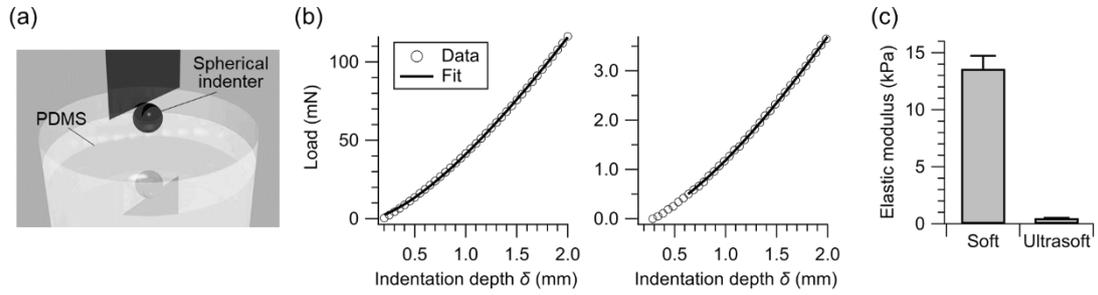

Figure 1

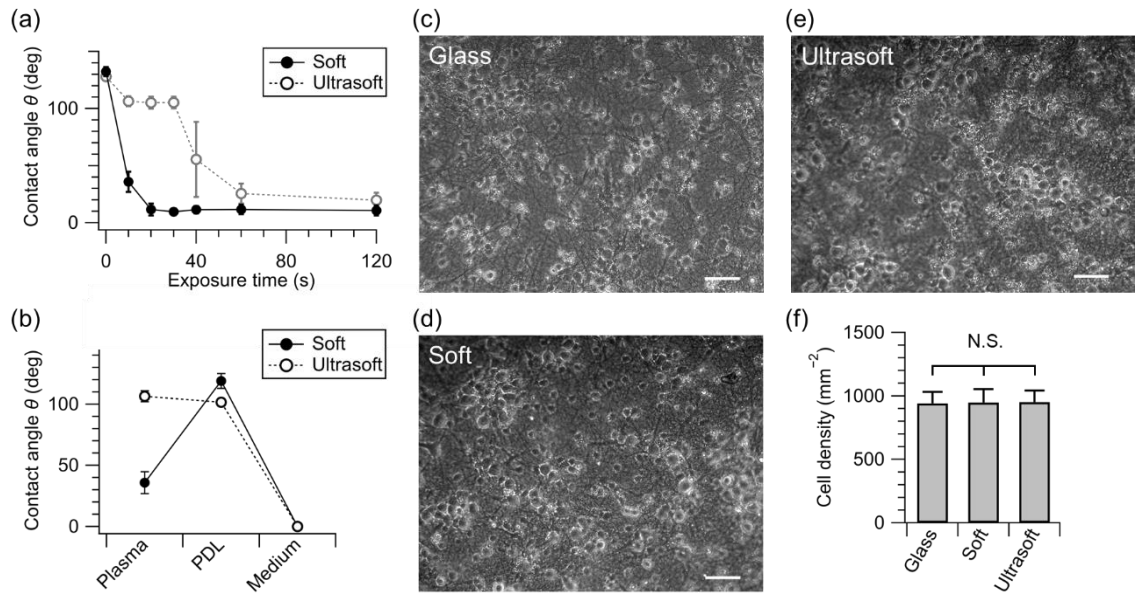

Figure 2

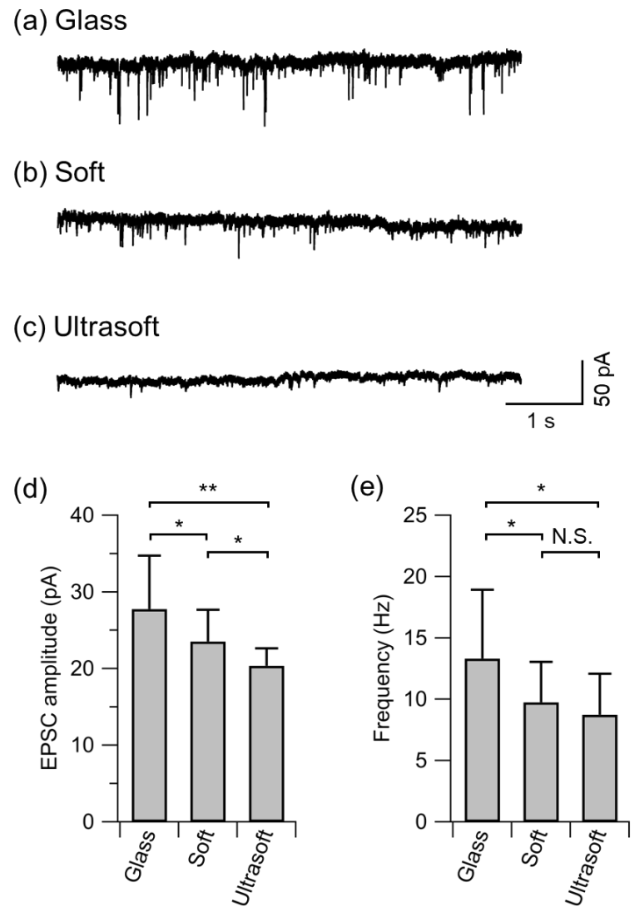

Figure 3

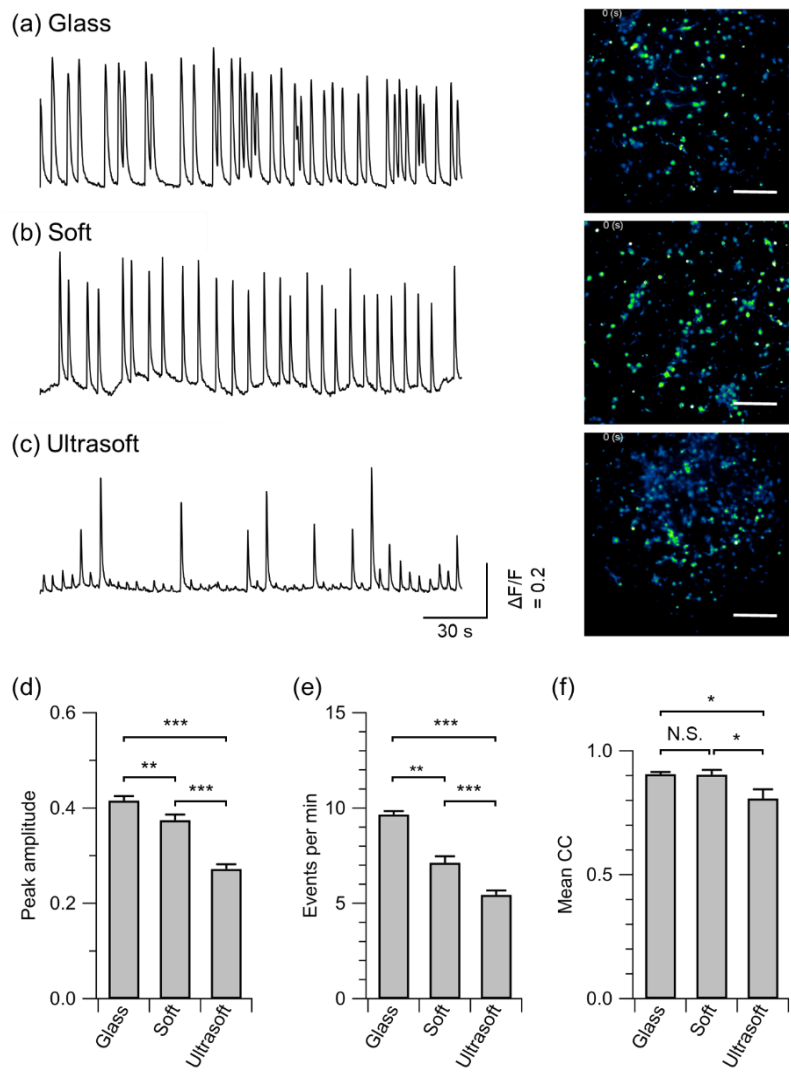

Figure 4

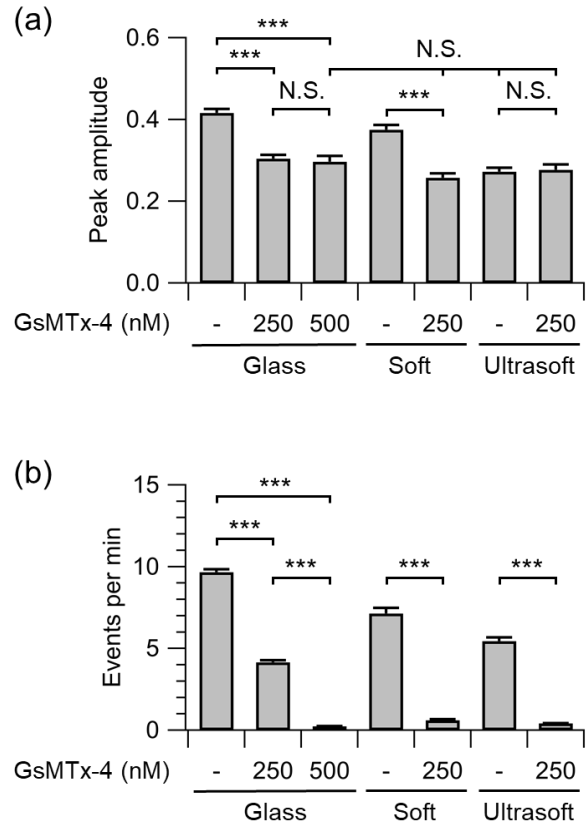

Figure 5

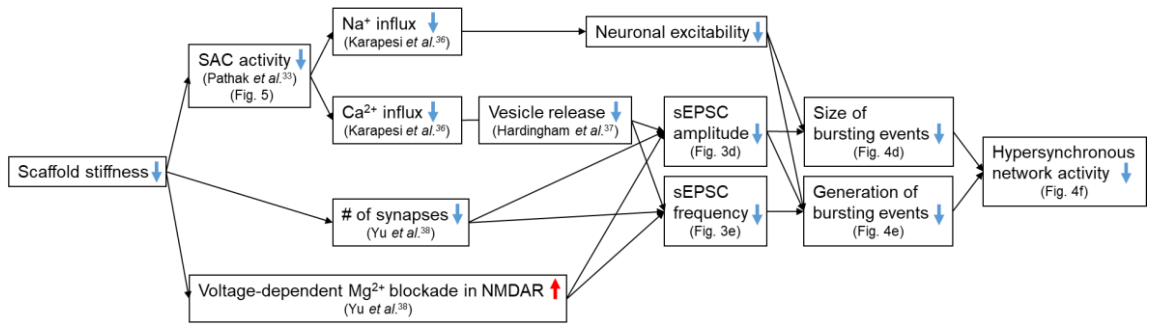

Figure 6

**Figure captions**

**Fig. 1.** Mechanical properties of PDMS. (a) Schematic illustration of the spherical indentation apparatus. (b) Load-displacement curves for soft (*left*) and ultrasoft (*right*) PDMS. Open circles represent the measured data, and the solid curve the fit with Eq. (1) ($r = 0.9999$ for both samples). For the data points, every 50th point is plotted for clarity. (d) Measured elastic moduli of soft and ultrasoft PDMS. Error bars, S.D.

**Fig. 2** Culturing primary neurons on PDMS. (a) Change in water contact angles of soft and ultrasoft PDMS upon exposure to air plasma. (b) Water contact angles measured after plasma irradiation for 10 s, after coating with PDL, and after immersion in the plating medium overnight. The surfaces of both samples were superhydrophilic after the immersion in the plating medium, and thus the data are plotted as 0°. (c–e) Primary cortical neurons cultured on (c) glass, (d) soft, and (e) ultrasoft scaffolds. Scale bars, 50 μm. (f) Average cell densities on the glass, soft, and ultrasoft substrates. Error bars, S.D.

**Fig. 3.** Effects of elastic modulus on sEPSC. (a–c) Representative recordings of spontaneous EPSCs on (a) glass, (b) soft, and (c) ultrasoft scaffolds. (d and e) The mean values of the amplitude (d) and frequency (e) of sEPSCs on respective surfaces. Error bars, S.D. * $p < 0.05$; ** $p < 0.01$.

**Fig. 4**. Impact of substrate stiffness on network activity of cultured cortical neurons. (a–c) Fluorescence intensity traces of representative neurons on (a) glass, (b) soft, and (c) ultrasoft scaffolds. Fluorescence micrographs are shown on the right. Scale bars, 100 μm. (d and e) Average peak amplitudes (d) and frequency of bursting events (e) on respective substrates. (f)

Mean correlation coefficient (mean CC) of neural activity on respective substrates. Error bars, S.E.M. * $p < 0.05$; ** $p < 0.01$; *** $p < 0.001$.

**Fig. 5**. Impact of the pharmacological blockade of SAC on neuronal network activity. (a and b) Average peak amplitudes (a) and rate of bursting events (b) at various concentrations of GsMTx-4 on respective substrates. Error bars, S.E.M. * $p < 0.05$; ** $p < 0.01$; *** $p < 0.001$.

**Fig. 6**. Diagram summarizing the present findings and the mechanisms underlying the suppression of hypersynchronous neuronal network activity on soft scaffolds.